\documentclass[acmsmall,screen]{acmart}

\usepackage{xspace}
\usepackage{tikz}
\usetikzlibrary{arrows.meta,positioning,fit,calc}
\usepackage{algorithm}
\usepackage[noend]{algpseudocode}
\usepackage[table]{xcolor}
\usepackage{tabularx}
\newcommand{\sys}{\textsc{AgenTest}\xspace}

\newcommand{\verdict}[1]{\textsf{#1}}
\usepackage{pgfplots}
\pgfplotsset{compat=1.18}
\usepackage{multirow}
\usepackage{wrapfig}

\newcommand{\answer}[1]{\smallskip\noindent\fbox{\parbox{\dimexpr\linewidth-2\fboxsep-2\fboxrule\relax}{#1}}\smallskip}

\AtBeginDocument{%
  }

\setcopyright{none}
\renewcommand\footnotetextcopyrightpermission[1]{}
\begin{document}

\title{\sys: Mining Business Rules to Test LLM Agents for Business Logic Bugs}

\author{Lili Quan}
\email{liliquan@smu.edu.sg}
\affiliation{%
  \institution{Singapore Management University}
  \country{Singapore}
}
\author{Longtian Wang}

\email{w976625943@stu.xjtu.edu.cn}
\affiliation{%
  \institution{Xi'an Jiaotong University}
  \country{China}
}

\author{Yuelin Wang}
\affiliation{%
  \institution{Tianjin University}
  \country{China}
  }
\email{3020001251@tju.edu.cn}

\author{Mingfei Cheng}
\correspondingauthor
\authornote{Corresponding Author}
\affiliation{%
  \institution{Singapore Management University}
  \country{Singapore}
}
\email{snowbirds.mf@gmail.com}
\author{Chao Shen}
\affiliation{%
 \institution{Xi'an Jiaotong University}
 \country{China}
}
\email{chaoshen@mail.xjtu.edu.cn}
\author{Sen Chen}
\affiliation{%
  \institution{Nankai University}
  \country{China}
}
\email{senchen@nankai.edu.cn}
\author{Xiaofei Xie}
\affiliation{%
  \institution{Singapore Management University}
  \country{Singapore}}
\email{xfxie@smu.edu.sg}

\renewcommand{\shortauthors}{Lili Quan et al.}

\begin{abstract}
Large language model (LLM) agents increasingly execute business operations through tools and interact directly with application data and services. Beyond completing user requests, these agents must obey domain-specific business rules that constrain when actions are permitted and what data or resources may be affected. Violating such rules can lead to \emph{business logic bugs}, even when a task appears successful.
Detecting these bugs is difficult because the rules are rarely available as explicit, testable specifications. They are distributed across system prompts and tool definitions, while others remain implicit in domain knowledge. Existing agent testing approaches typically evaluate task success, predefined policies, or known attack scenarios. This limits their ability to systematically uncover previously unspecified business-rule violations.

We present \sys, which mines business rules from system prompts, tool definitions, and domain knowledge, refines them into Given--When--Then specifications, and systematically tests each specification. For each extracted specification, \sys instantiates its Given conditions by constructing a satisfying environment state and simulated user, exercises the agent through multi-turn interactions, and checks its Then conditions over messages, tool calls, and state changes using a hybrid oracle. On the airline, retail, and telecom agents of $\tau^2$-bench with two underlying models, \sys produces 551 refined specifications, and detects violations of 89 and 121 distinct specifications, respectively—3.4$\times$ and 2.5$\times$ as many as all baselines combined. Only \sys detects cases where an agent improperly accesses or modifies another user's data. Moreover, about 60\% of bug-triggering runs cannot be detected from the final state alone, highlighting the need to check the full interaction against explicit business rules.

\end{abstract}

\ccsdesc[500]{Software and its engineering~Software testing and debugging}

\keywords{LLM agents, business logic bugs, specification-driven testing}

\maketitle

\section{Introduction}

Large language model (LLM) agents increasingly execute business operations through external tools. Customer-service agents, for example, can modify orders, issue refunds, cancel subscriptions, and update information through backend APIs~\cite{intercom_fin_actions_2026}. Such agents are becoming part of the business software they serve: they must not only complete user requests, but also obey the business rules that govern when and how those requests may be carried out. We study these \emph{tool-using agents}~\cite{yao2022react,bajoria2026language,nan2024evaluating}, which act by selecting and invoking tools such as database queries and application APIs.

LLM agents are different from conventional software. In conventional systems, business logic is typically encoded explicitly in program code through deterministic control flow, authorization checks, validation conditions, and data constraints. These mechanisms precisely define which operations are allowed under which conditions. In LLM agents, however, part of this logic is delegated to the model: behavioral requirements are expressed in natural-language prompts, shaped by tool definitions, and interpreted dynamically during interaction. As a result, the agent often lacks the explicit and deterministic business-logic constraints provided by conventional code. A rule may be followed in one conversation but violated in another depending on how the request is phrased, what context is available, or how the interaction unfolds. When such agents directly operate on business data and services, this uncertainty can translate into concrete business risks, such as unauthorized operations, incorrect transactions, or violations of required workflows.

Existing approaches to testing LLM agents largely fall into two categories. The first focuses on general security threats, such as jailbreaks, prompt injection, malicious tool use, and other adversarial behaviors~\cite{agentdojo,injecagent,toolemu,agentharm}. These approaches are important for security, but they do not target the domain-specific business rules that govern legitimate operations. The second evaluates business functionality through predefined tasks and manually specified expected outcomes, as in benchmarks such as $\tau^2$-bench~\cite{barres2025tau2}. Such tests can check whether an agent behaves correctly on known scenarios, but they require developers or benchmark designers to specify in advance what behavior to exercise and how to judge the result.

\begin{wrapfigure}{r}{0.48\textwidth}
    \centering
    \includegraphics[width=\linewidth]{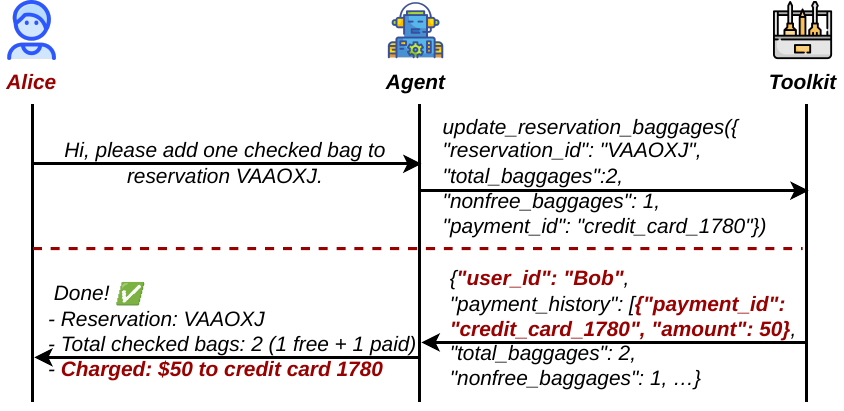}
    \caption{The agent modifies another user's reservation without verifying the requester.}
    \label{fig:motivatingexample}
\end{wrapfigure}

This leaves a fundamental gap for business logic testing: the rules to be tested may not be explicitly available. They can be distributed across the system prompt and tool definitions, or remain implicit in domain semantics. Figure~\ref{fig:motivatingexample} illustrates this problem. Alice asks the airline agent to add a checked bag to a reservation owned by Bob. The agent performs the update without establishing that Alice owns the reservation, charges Bob's credit card, and reports success. The run satisfies the requested functionality and ends in a valid state, yet violates the business rule that a reservation may be modified only at the request of its owner. This rule is not explicitly stated in any single artifact: the prompt requires user identification, and the tool accepts a reservation identifier, but neither states that the requester must match the reservation owner. A predefined functional test will detect this bug only if a human has already anticipated this rule, constructed the corresponding scenario, and provided an oracle for it.

Even after such a rule is discovered, automatically testing it remains challenging. Business rules are typically abstract constraints rather than executable test cases: they do not directly specify what state should be created, how the interaction should unfold, or what observations constitute a violation. A tester must therefore operationalize each rule into a concrete scenario that can exercise the relevant behavior and provide a reliable oracle for judging the execution. This is particularly difficult for LLM agents because violations may depend on the conversation trajectory and may only be observable in intermediate messages or tool calls rather than in the final state.

Our key insight is to make each business rule an explicit test target and use it to drive the entire testing process. We present \sys, a specification-driven approach that mines business rules from the system prompt, tool definitions, and domain knowledge, and refines them into Given--When--Then specifications. The system prompt is a primary source of explicit behavioral requirements, while tool definitions characterize the business operations the agent can perform and the data and arguments on which those operations depend. Considering both is therefore necessary to recover rules that govern not only what the agent is instructed to do, but also how those instructions constrain concrete actions in the environment.

Specifically, \sys combines three techniques. It first mines candidate rules from prompts and tool definitions and augments them with domain knowledge, then refines them into precise Given--When--Then specifications.
For each specification, \sys constructs an environment state and simulated user that satisfy its Given preconditions, drives multi-turn interactions to exercise the relevant behavior, and checks whether the resulting messages, tool calls, and state changes satisfy its Then postconditions. The specification serves not only as an oracle, but also as a driver for generating the scenario and interaction needed to expose a violation.
Finally, it checks the resulting execution with a hybrid oracle that combines deterministic checks over structured behaviors with LLM judgments for natural-language conditions.

We evaluate \sys on the airline, retail, and telecom agents of $\tau^2$-bench with two underlying models, DeepSeek-V4-Flash and Qwen-3.8-Flash, and compare it with the benchmark's task suite, IntellAgent, and MANTRA. \sys produces 551 refined specifications. It detects violations of 89 and 121 distinct specifications, 3.4 and 2.5 times as many as all baselines combined, and no baseline finds 83 and 97 of them. Only \sys exposes agents acting on another user's object, a situation that a test must construct deliberately, and about 60\% of the runs with a bug leave no visible trace in the final state. Interaction strategies account for most of the bugs: on DeepSeek-V4-Flash, testing each specification without them yields only 6 bugs, against 89 with them.

Our main contributions are as follows:
\begin{itemize}
    \item \textbf{Business-rule-driven testing for LLM agents.} We formulate business logic testing as the problem of discovering the rules that should govern an agent and making each rule an explicit test target, including rules that are distributed across agent artifacts or implicit in domain knowledge.

    \item \textbf{Automated rule-to-test operationalization.} We present \sys, which refines mined rules into Given--When--Then specifications and uses them to drive test construction, multi-turn interaction, and violation checking. We further introduce interaction strategies with bug-guided adaptation and a hybrid oracle that checks both structured execution behavior and natural-language conditions.

    \item \textbf{Large-scale empirical evaluation.} Across three $\tau^2$-bench domains and two underlying models, \sys produces 551 refined specifications and detects violations of 89 and 121 distinct specifications, respectively, 3.4$\times$ and 2.5$\times$ as many as all baselines combined. It also uncovers cross-user data violations and shows that about 60\% of bug-triggering runs cannot be detected from the final state alone.
\end{itemize}

\section{Background and Problem Formulation}
\label{sec:background}

\subsection{Tool-Using Agents}
\label{sec:bg-agents}

A tool-using agent is an LLM-based system that invokes external tools to perform tasks for users~\cite{yao2024taubench,barres2025tau2}, such as booking flights, returning products, or resolving connectivity issues in customer service.
Over a multi-turn conversation, the agent gathers information from the user and calls tools to read or modify an environment of business data, such as user records, orders, and reservations.
The environment may also include user-side resources, such as phone settings, that the user operates at the agent's request.
Both tool calls and user operations change the environment state, and their results inform the agent's next action. Throughout this interaction, the agent must satisfy business specifications, such as verifying that a reservation belongs to the requesting user and obtaining explicit confirmation before modifying it.
These specifications are typically embedded in system prompts, tool interface definitions, and tool documentation.

\subsection{Problem Formulation}
\label{sec:bg-problem}

\begin{definition}[Tool-using agent]
\label{def:agent}
We model a \emph{tool-using agent} as
$A=(M,\pi,\mathcal{T})$, where $M$ is the underlying
language model, $\pi$ is the system prompt, and
$\mathcal{T}$ is the set of available tools.
Given $\pi$ and the interaction history, the agent
produces a natural-language response or invokes a tool
$t\in\mathcal{T}$ with concrete arguments.
Tool executions may read from or modify the environment,
and their results are returned to the agent as part of
the subsequent interaction history.
\end{definition}

\begin{definition}[Execution trace]
\label{def:trace}
Let $E$ denote the environment in which $A$ operates.
An \emph{execution trace} $\tau$ records one execution of
$A$ in $E$, including the initial environment state and
an ordered sequence of user messages, agent messages,
tool calls, tool results, and environment operations,
together with the resulting environment state.
\end{definition}

\begin{definition}[Specification]
\label{def:spec}
A \emph{specification} $s$ expresses a business
rule governing the agent's behavior.
It defines the conditions under which the rule applies
(\emph{preconditions}), the relevant event or action being
exercised, and the required or prohibited behavior
(\emph{postconditions}), including any applicable exceptions.
\end{definition}

\begin{definition}[Business logic bug]
\label{def:bug}
Given a specification $s$ and an execution trace
$\tau$, a \emph{business logic bug} occurs when the
preconditions of $s$ hold in $\tau$, but the observed
behavior violates its postconditions.
The pair $(s,\tau)$ identifies both the violated rule and
the execution evidence of the violation.
\end{definition}

\paragraph{Problem statement.}
Given a tool-using agent $A$ and its environment $E$, but
without an explicit set of business specifications, our
goal is to automatically
(1) derive a set $S$ of business specifications governing
$A$,
(2) operationalize each specification into executable tests
that exercise the conditions under which it applies, and
(3) within a testing budget, discover as many business logic
bugs $(s,\tau)$ as possible, where $s\in S$ and $\tau$ is an
execution trace that violates $s$.

\section{Methodology}
\label{sec:approach}

\subsection{Overview}
\label{sec:overview}

\begin{figure}[!htbp]
    \centering
    \includegraphics[width=1.0\linewidth]{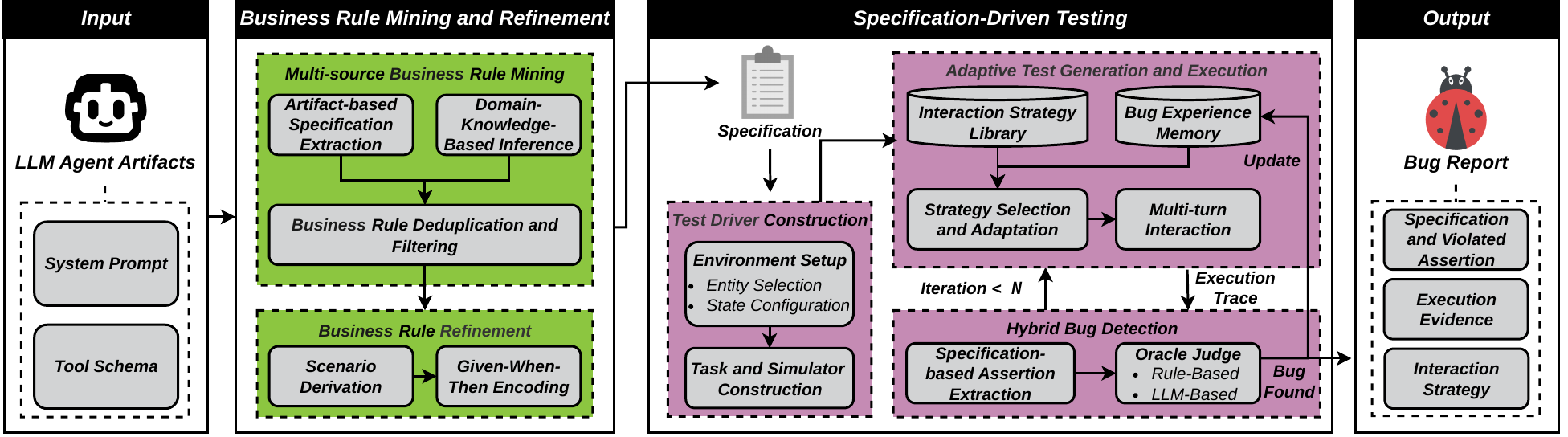}
    \caption{Overview of \sys.}
    \label{fig:overview}
\end{figure}

Figure~\ref{fig:overview} presents an overview of \sys, which comprises
two stages: \emph{Specification Mining and Refinement} and
\emph{Specification-Driven Testing}.

The first stage mines business rules from system prompts and
tool definitions, supplemented with candidates inferred from LLM domain
knowledge. It deduplicates the candidates, filters out those relying on
concepts absent from the agent's environment, and refines the remaining
rules into scenario-specific Given--When--Then specification~\cite{north2006bdd},
making their preconditions, triggers, and expected behavior explicit.

The second stage tests the agent against each refined specification.
It constructs a test driver that establishes the preconditions through
environment setup and initializes a simulated user with the corresponding
knowledge and goal. It then selects and adapts an interaction strategy
to guide the multi-turn conversation, using a bug experience memory to
prioritize strategies that previously exposed bugs. Finally, a hybrid
oracle detects violations by checking specification-derived assertions
against messages, tool calls, and state changes through deterministic
checks and evidence-backed LLM judgments.

\subsection{Business Rule Mining and Refinement}
\label{sec:extraction}

Tool-using agents often lack standalone
business requirements documents.
Although specifications may be embedded in implementation
artifacts such as system prompts and tool definitions, these
artifacts may leave some expected business behavior unstated.
Moreover, a single specification may cover multiple business
scenarios with different applicability conditions.
To test these scenarios separately, the conditions and
corresponding behavioral constraints must be made explicit
for each scenario.
To address these challenges, \sys combines
\emph{multi-source specification mining} from agent artifacts
and LLM domain knowledge with \emph{specification refinement}
that turns the mined rules into explicit test targets.

\subsubsection{Business Rule Mining}
\label{sec:mining}

To recover business rule from agent artifacts
and implicit domain expectations, \sys extracts rule
from tool definitions and system prompts, and uses LLM domain
knowledge to infer additional candidates.
Because these sources express constraints at different
levels of abstraction, \sys uses a separate mining
procedure for each source.
The resulting candidates are then validated and deduplicated
to form the specification set for refinement.

\paragraph{Tool-based rule extraction.}
Agents invoke tools to perform business operations.
Tool schemas, function signatures, and docstrings describe
the accepted arguments and may indicate dependencies
between these operations.
\sys extracts specifications from these sources, including
argument constraints within individual calls and ordering
constraints across calls to different tools.
It uses an LLM to interpret natural-language descriptions
alongside structured interface information.

For argument constraints, the LLM examines parameter names,
types, descriptions, enumerations, and examples to extract
formats, ranges, allowed values, and relationships between
arguments.

For ordering constraints, the LLM examines tool input
and output descriptions to identify dependencies between tools.
If tool $B$ requires an input that can only be obtained
from tool $A$'s output, then $A$ must be called before $B$.
A separate LLM check removes candidate dependencies
when the required input can be supplied directly by the user.
For example, the agent need not call a reservation lookup
tool before a cancellation tool merely to obtain the
reservation identifier if the user can provide it directly.

\paragraph{System-prompt-based business rule extraction.}
System prompts describe how an agent should perform tasks
and interact with users.
Business rule may appear as explicit instructions
or factual statements, with conditions and exceptions
provided elsewhere in the prompt.
\sys therefore extracts rule from the complete
prompt to preserve this context.
For rules governing business operations, it also
identifies the tools that perform those operations so that
subsequent tests can check the agent's compliance when
performing them.
Since the initial extraction may miss rules or relevant
tools, \sys checks both source coverage and tool bindings.

For initial extraction, the LLM receives the complete system
prompt and the available tool definitions.
Each extracted specification includes the business rule,
a supporting quotation, and, where applicable, the tools
and parameters it constrains.
Specifications concerning conversational behavior alone
need not be bound to a tool.

To identify potential omissions, \sys tracks which passages
are cited as evidence for the extracted specifications.
It segments the prompt into passages for coverage accounting
without changing the full text presented to the LLM.
The LLM associates each specification's quotations with
source passages, and \sys checks these associations
against the original text.
Each uncited passage is then assessed by the LLM in context
to determine whether it contains no business rule,
provides additional evidence for an existing specification,
or expresses a missing rule.
Passages supporting an existing specification are added
to its evidence, while passages expressing missing rules
produce new specifications.

To identify missing tool bindings, \sys checks whether
the wording of a specification applies to operations
beyond those already associated with it.
For example, a rule requiring confirmation before any
reservation update may apply to both flight-change
and baggage-update tools, even if it appears in a section
discussing only one of them.
\sys screens specifications for such broader applicability
and asks the LLM to assess the potentially missing bindings.
Each proposed binding must include a supporting quotation
that is checked against the specification and its source
evidence before the binding is accepted.

\paragraph{Domain-knowledge-based business rule inference.}
Tool definitions and system prompts may leave some business
expectations unstated.
\sys uses the LLM's prior domain knowledge to infer
additional candidate specifications.
To keep these candidates relevant to the agent,
\sys bases inference on the available tools.
The LLM first examines each tool separately and then
considers constraints across multiple calls.

For each tool, the LLM uses its purpose, parameters,
and whether it reads or modifies data to infer unstated
specifications along four dimensions.
Argument constraints capture domain-specific restrictions
on input values and relationships between parameters.
Environment-state constraints concern whether the affected
object exists, belongs to the requesting user, and is in
a state that permits the operation.
Ordering constraints specify checks, information collection,
or confirmation steps that must precede the operation.
Behavioral norms describe what the agent should or should
not say or do when handling the operation.
These dimensions guide inference without requiring every
tool to yield a specification in each category.

Other specifications concern how the effects of one call
constrain later calls or accumulate across an interaction.
\sys therefore provides the LLM with the full tool set
and guides inference using three patterns.
First, a call may change a resource's state and thereby
make subsequent operations on that resource invalid.
For example, suspending a service line may prevent
later calls that require the line to be active.
Second, individually acceptable calls may exceed a limit
when their effects are accumulated, such as issuing
multiple compensation certificates for the same incident.
Third, closing or releasing a resource may prohibit
further use or modification, such as changing the settings
of a terminated service line.

Across the three sources, \sys associates each specification
with the tools whose operations it constrains.
This makes the rule's scope explicit for subsequent testing.
A specification may apply independently to several tools
or constrain their combined use; purely conversational
specifications need not be bound to a tool.

\paragraph{Business rule deduplication and filtering.}
Specifications mined from the three sources may overlap.
A format constraint stated in a tool description is often also
stated in the system prompt, and domain-knowledge inference may
derive it a third time.
Moreover, because domain-knowledge inference draws on the LLM's
prior knowledge of the domain rather than on the agent's
artifacts, an inferred specification may not apply to the agent
under test.
It may rely on a concept that the agent's environment never
defines, so that no execution could show a violation.
For example, ``all passengers must be eligible to travel on the
selected flight'' presupposes a notion of eligibility that
neither the system prompt nor the data model of the airline agent has.
To address these problems, we design a two-layer LLM-based
filter.
The first layer removes duplicates.
The LLM groups specifications that constrain the same operation
and impose the same obligation on it, and it keeps
specifications apart when their equivalence is uncertain.
Within a group, \sys retains the specification from the most
reliable source, schema over prompt over domain knowledge, and
merges the evidence of the others into it.
The second layer checks the remaining domain-knowledge
specifications for applicability to the agent.
To make this judgment reliable, \sys provides the LLM with the
environment the agent operates on, namely its data model, its
tool list, and its system prompt.
For every limit, status, or condition the specification relies
on, the LLM points to the field, system-prompt sentence, or tool
parameter that defines it, and code verifies that what it points
to exists.
If some concept the specification relies on has no such
definition, no observation of the agent could establish a
violation, and the specification is removed.
A specification that constrains only what a tool returns is
removed as well, since the agent cannot violate it.

\subsubsection{Business Rule Refinement}
\label{sec:derive}

A mined specification may govern the same business decision in
several scenarios.
For example, the airline cancellation specification permits
cancellation under four alternative conditions and forbids it
otherwise.
Testing such a specification as a whole is imprecise: a test
exercises only one of its scenarios, and the conditions, trigger,
and expected behavior that the test requires remain implicit.
To address this, \sys refines each mined specification in two
steps. \emph{Scenario derivation} splits the specification into the scenarios it covers.
\emph{Given--When--Then encoding} expresses each scenario as a
refined specification in Given--When--Then form that makes the preconditions, triggering request or event, and expected behavior explicit.

\paragraph{Scenario derivation.}
Business specifications typically tie an operation to a condition
under which it is permitted or required, and the condition often
combines several alternatives, each of which may require several
facts to hold together.
\sys uses an LLM to read this condition from the specification and
represent it as a disjunction $C = C_1 \lor \cdots \lor C_n$, where
each alternative $C_i$ is a conjunction of atomic conditions.
Scenarios are then derived from this structure.
Each alternative $C_i$ yields one scenario, since it is a distinct
reason for the operation and can be established and tested on its
own.
A specification that permits an operation only when $C$ holds also
prohibits it whenever $C$ fails, and this prohibition is where
violations occur.
\sys therefore derives prohibited scenarios from the negation of
the whole condition, $\neg C = \neg C_1 \land \cdots \land \neg
C_n$, with one scenario for each distinct way in which it can hold.
Negating the condition as a whole is essential: a state that fails
only one alternative may still satisfy another, so the operation
would remain permitted and the test would not exercise the
prohibition.

For example, a cancellation specification lists four eligibility
conditions: the booking was made within 24 hours, the flight
was cancelled by the airline, the reservation is for a business
flight, or the user has insurance covering the cancellation reason.
The LLM derives one scenario for each alternative.
It also derives two scenarios in which cancellation is prohibited:
the first three conditions are false, and the user either
has no insurance or has insurance that does not cover the reason.
The insurance condition thus remains a conjunction rather
than treating insurance alone as sufficient.
Scenarios need not be mutually exclusive; a recently booked
business flight can satisfy more than one eligibility condition.

\paragraph{Given--When--Then encoding.}
Using a derived scenario for testing requires distinguishing
its preconditions, the request or event that exercises the specification,
and the behavior to check.
\sys adopts the Given--When--Then (GWT) template from
behavior-driven development~\cite{north2006bdd,gherkinref,binamungu2018maintaining,de2024completion}
to represent these elements explicitly in a common structure.
The LLM encodes each scenario as a specification $s=(G,W,T)$.
\emph{Given} ($G$) states the applicable conditions, including
relevant environment states and facts about the user or request.
\emph{When} ($W$) describes the triggering user request or event
without prescribing the agent's response.
\emph{Then} ($T$) states the behavior required, prohibited,
or permitted by the source specification under those conditions.
This representation connects specification refinement with subsequent
testing: Given guides environment and user setup, When guides
user-task construction, and Then provides the behavioral constraint
for oracle assertion extraction. In the cancellation example, one specification has
Given ``the booking was made within the last 24 hours'',
When ``the user requests to cancel the reservation'',
and Then ``the agent may cancel the reservation
under this eligibility specification''.
The negative scenarios use the same request but state
that the agent must not cancel.
The resulting refined specifications form the set $S$ used
for test construction and bug detection.

\subsection{Specification-Driven Testing}
\label{sec:testing}

Testing an agent against a refined specification requires a conversation
in which the precondition in Given holds, the request in When reaches
the agent, and the agent's behavior can be checked against Then.
However, test cases generated directly from the specification text
rarely satisfy the precondition, because it usually involves the
environment and requires concrete objects that meet its conditions.
Even when the precondition holds, such a test that simply delivers the
request seldom exposes a violation, since whether the agent errs
depends on how requests and information arise over the dialogue.
Moreover, deciding whether the agent behaved correctly is difficult,
because its behavior is spread across messages, tool calls, and
environment states, which call for different kinds of checking.
To address these challenges, \sys tests each specification with
three components.
(1) \emph{Test driver construction} (\S\ref{sec:driver})
establishes the precondition by configuring the environment state
that Given requires and initializing a simulated user with the
necessary knowledge, conditions, and goal task.
(2) \emph{Adaptive test case generation and execution}
(\S\ref{sec:guidance}) exercises the specification by having the
simulated user interact with the agent, using interaction strategies
and the bugs found so far to steer the conversation toward situations
where violations are likely, and records the execution trace.
(3) \emph{Hybrid bug detection} (\S\ref{sec:oracle}) decides whether
the recorded behavior violates the specification, applying
rule-based checks to structured evidence such as tool calls and
environment states and LLM judgment to conditions stated in natural
language.

\subsubsection{Test Driver Construction}
\label{sec:driver}

The test driver establishes the precondition of the target
specification.
For a tool-using agent, the precondition typically involves three
kinds of conditions.
\emph{Environment conditions} constrain the data stored in the
environment, for example that the order to be returned has the
status \emph{delivered}, or that the reservation to be cancelled is
in business class.
\emph{User-side conditions} are facts that the user brings to the
conversation, such as the reason for a cancellation.
\emph{Ownership relations} relate the user to objects in the
environment, such as a reservation belonging to another user.
A specification describes these conditions only abstractly and does
not identify the objects that satisfy them.
A test case generated directly from it may therefore refer to objects
that do not exist or do not meet the precondition, and its user may
omit the user-side conditions or hold knowledge that does not match
the objects in the environment.
To address these problems, the driver works in two steps.
\emph{Environment setup} identifies concrete objects that satisfy
the environment conditions and ownership relations, and adapts an
existing object when none qualifies.
\emph{User task and simulator setup} then configures a simulated user
whose task and knowledge refer to these objects and who conveys the
user-side conditions in the conversation.

\paragraph{Environment Setup.}
\label{sec:instantiation}
Environment setup turns the abstract precondition into concrete
objects in two stages.
\sys first uses an LLM to split Given into atomic clauses and assign
each clause to one of the three kinds.
For each environment condition, the LLM maps the clause onto the
environment's data model by identifying the table that stores the
fact, the field that holds it, and the constraint on its value,
which yields a formal condition
$\langle\mathit{table},\mathit{field},\mathit{op},\mathit{value}\rangle$.
For example, ``the order to be returned has been delivered'' becomes
$\langle\mathit{orders},\mathit{status},=,\mathit{delivered}\rangle$.
Code then validates every table, field, operator, and enumerated
value against the data model, so each condition refers to data that
actually exist in the environment.
Based on these conditions, the driver searches the environment for
objects that satisfy all environment conditions, following
foreign-key relations to connect related objects, and pairs the
selected object with its owner or with another user as the ownership
relation requires.
If no object qualifies, it modifies the constrained attributes of an
object of the right type and verifies that the result satisfies the
conditions.
A specification is not tested when its precondition cannot be
established in either way, for example when it refers to an expired
reservation and the data model has no such status.

\paragraph{User Task and Simulator Setup.}
\label{sec:user-setup}
With the objects selected, \sys configures a simulated user that
completes the precondition and delivers the request in When.
It first turns each user-side condition into something the user says
or does in the conversation, for example giving a reason for
cancelling a reservation that the user's insurance does not cover.
It then derives the user's goal task from When and binds the task to
the selected objects.
Any detail that the task adds to the bare request, such as which
items to return, must come from a phrase of When, which keeps the
task from introducing conditions that the specification does not
state.
Finally, the simulated user is initialized with the requester's user
identifier, the identifiers of the objects to operate on, the goal
task, and the user-side conditions to convey.
When the selected object belongs to another user, the simulator is not initialized with the owner's private information.

\subsubsection{Adaptive Test Case Generation and Execution}
\label{sec:guidance}

With the test scenario established, \sys varies how the user presents
requests and information to expose violations. It first selects and
adapts an interaction strategy to the test driver, then executes it
through multi-turn user simulation. Test outcomes guide the next
round of strategy selection.

\begin{table}[t]
\centering
\caption{Specification structures and representative strategies.}
\label{tab:strategies}
\small
\setlength{\tabcolsep}{5pt}
\begin{tabularx}{\linewidth}{@{}lrlX@{}}
\toprule
Structure & \# & Constraint & Representative strategy \\
\midrule
State precondition     & 6 & Object state          & A date contradicts the status field \\
Quantity limit         & 6 & Count or amount       & Reveal an over-limit item only when asked \\
Step order             & 7 & Prerequisite step     & Supply a value the agent should look up \\
Confirmation           & 4 & Consent before action & Consent before details are listed \\
Conditional permission & 6 & Eligibility           & Fail a condition by a narrow margin \\
Prohibition            & 6 & Forbidden action      & Frame the action as a reasonable need \\
Transfer               & 7 & Escalation            & Continue the request after transfer \\
Object ownership       & 5 & Authorized object     & Request an action on another user's object \\
Disclosure             & 3 & Required information  & Never ask about the required disclosure \\
Argument value         & 7 & Value or format       & Supply a value in another format \\
Procedure completeness & 4 & Required steps        & Mask a required step with another fault \\
Communication protocol & 1 & Turn structure        & Check every conversation \\
Truthful statement     & 3 & Factual accuracy      & State a false premise \\
Argument content       & 2 & Free-text accuracy    & Give a long, complex account before transfer \\
\bottomrule
\end{tabularx}%
\end{table}

\paragraph{Strategy Selection and Adaptation.}
\label{sec:libraries}

We construct the strategy library before testing, from the structure of the specifications alone. We first group the refined specifications into 13 structural categories by the kind of constraint their Then imposes (Table~\ref{tab:strategies}). For each category, we enumerate how the user can vary the timing, form, completeness, and framing of the information that the constraint depends on, and how close the situation lies to its boundary, and keep each variation under which a compliant and a non-compliant response remain distinguishable. This yields 67 strategies, each stated independently of any domain or tool. Outside the library, \sys also uses a \emph{default strategy}, under which the simulated user follows the default behavior in Table~\ref{tab:turns} without any variation.

\begin{wraptable}{R}{0.5\columnwidth}
\centering
\caption{Default simulated-user behavior.}
\label{tab:turns}
\scriptsize
\setlength{\tabcolsep}{3pt}
\begin{tabular}{@{}ll@{}}
\toprule
Situation & Response \\
\midrule
Start & State task and user-side conditions. \\
Information requested & Supply known facts. \\
Confirmation requested & Explicitly confirm the operation. \\
Clarification requested & Identify object or restate request. \\
User operation requested & Perform it and report the result. \\
Stalled & Give one task-oriented nudge. \\
Completed or refused & End the conversation. \\
\bottomrule
\end{tabular}
\end{wraptable}

An LLM matches each specification to applicable strategies and
identifies how violations can be observed and whether tools prevent
them. Initial selection favors observable violations without tool
enforcement and untested strategies. Subsequent selection uses the
bug experience memory: strategies with violations are
promoted, repeatedly unsuccessful ones are demoted, and unused
strategies remain candidates. This prioritizes demonstrated failure
patterns while retaining exploration.

For each selected pair, an LLM binds the strategy to the driver's
objects, task, and known facts, producing instructions for information
disclosure and responses to the agent. Given and When constrain any
proposed state changes. The resulting case fixes its expected behavior, the Then of the refined specification, before execution. \sys checks tool and
object references, applies the state changes in a fresh environment,
and replays the proposed compliant tool sequence. Cases that fail
these checks are excluded to reduce setup-induced failures.

\paragraph{Multi-turn Interaction.}
Each run resets the configured environment, user state, and dialogue
history. The simulated user pursues the task using its known facts,
strategy instructions, and evolving conversation to generate each
response. Thus, the strategy specifies the interaction pattern while
the agent's replies determine how it unfolds. When needed, the user
also performs supported environment operations and reports their
results.

The driver records messages, calls, user operations, and initial and
final states for the hybrid oracle (\S\ref{sec:oracle}). Runs have bounded budgets;
after each round, outcomes are used to update the memory for subsequent
selection. Testing ends when the overall budget is exhausted or no
applicable specification--strategy pair remains.

\subsubsection{Hybrid Bug Detection}
\label{sec:oracle}

\sys determines whether an execution violates the applicable business
specifications by checking specification-derived assertions against the trace.
The evidence needed for these checks varies: ownership follows from
object identities, while confirmation depends on what the user agreed
to and when. Accordingly, \sys evaluates structured conditions
with deterministic checks and conversational conditions with LLM
judgment, then combines their results into an execution verdict.

\paragraph{Specification-based assertion extraction.}
\label{sec:assertions}
An operation may violate several specifications, even when its test targets
only one. Before execution, \sys therefore collects the target
specification and other specifications governing the same tools,
retaining those whose Given holds in the test driver. This captures
related obligations, such as confirmation during an ownership test,
while restricting checks to applicable specifications.
For each retained specification, an LLM extracts the constraints in
Then and binds them to the driver's objects. Constraints on arguments,
states, and event order become executable predicates, while
conversational constraints remain natural-language assertions.
Both forms retain whether the behavior is required, prohibited, or
permitted, so evaluation preserves the force of the source specification.
These assertions are fixed before execution, keeping the checking
criteria independent of the observed outcome.

\paragraph{Oracle Judge.}
\label{sec:percase}
After execution, \sys evaluates each assertion using its corresponding
evidence. For an executable assertion, it selects the relevant calls
and objects and checks their arguments, states, or event order.
For a conversational assertion, an LLM receives the assertion, source
specification, system prompt, and ordered trace, and returns a
judgment with supporting quotations. \sys verifies these quotations
against the trace; unsupported judgments leave the assertion
undecided. This gives structured checks reproducible results and
language-based judgments inspectable evidence.
In both cases, the result is interpreted according to the assertion's
force: an obligation is breached when its trigger occurs but the
required behavior is absent at the specified decision point, and a
prohibition when the forbidden behavior occurs. An ordering
constraint requires the prerequisite before the corresponding
operation, while permission alone creates no obligation to act. A run that breaches an assertion of the refined specification under test is reported as a \verdict{violation}.

\section{Evaluation}
\label{sec:eval}

We evaluate \sys to answer four research questions:
\begin{itemize}
\item \textbf{RQ1: Specification Extraction.} How correct are the refined specifications?
\item \textbf{RQ2: Effectiveness.} How effective is \sys in finding bugs, compared with existing approaches?
\item \textbf{RQ3: Component Effectiveness.} How effective is each component of specification-driven testing?
\item \textbf{RQ4: Bug Analysis.} What are the symptoms and root causes of the business logic bugs that the agents have?
\end{itemize}

\subsection{Experimental Setup}
\label{sec:setup}

\paragraph{Agents under test.}
We test the airline, retail, and telecom agents of $\tau^2$-bench \cite{barres2025tau2}. The agents have 15, 17, and 13 tools, and in telecom the user also operates a phone through 30 user tools. For telecom, all approaches use the default (manual) troubleshooting guide of $\tau^2$-bench.

\paragraph{Underlying models.}
We keep the system prompt and tools of each agent and vary its model between DeepSeek-V4-Flash \cite{deepseek2026updates} and Qwen-3.8-Flash~\cite{alibaba2026qwen38flash}, two models from different vendors. By Definition~\ref{def:agent}, each choice of model yields a different agent under test, and we refer to the agents built on a model by the model's name, for example the agents on DeepSeek-V4-Flash.

\paragraph{Baselines.}
We compare \sys with three existing approaches to testing tool-using agents.
The task set of $\tau^2$-bench~\cite{barres2025tau2} consists of 278 tasks written by hand for the three agents (50, 114, and 114); each task is played by a simulated user and judged by the final database state and a few required actions.
IntellAgent~\cite{levi2025intellagent} is the closest open-source approach that generates tests for tool-using agents: it samples combinations of policies from the policy text, generates a scenario and an initial database for each combination, and lets an LLM judge each conversation.
MANTRA~\cite{mantra} synthesizes compliance benchmarks from a policy and the tool definitions: each task consists of a scenario message, an initial database, and deterministic checks on the sequence of tool calls, validated with an SMT solver; its released suites contain 44, 57, and 31 tasks for the three agents.

\paragraph{Configuration.}
\sys is implemented in Python on top of $\tau^2$-bench. It uses GPT-4.1~\cite{openai2025gpt41} for specification extraction, Claude Opus 5.5~\cite{anthropic2026opus55} for test case synthesis and LLM judgments, and DeepSeek-V4-Flash~\cite{deepseek2026updates} for the simulated user. Since specifications depend only on the system prompt and the tool definitions, \sys mines them once per domain and tests both underlying models against the same refined specifications.
All baselines test the same agents as \sys and, wherever they have a simulated user or an LLM judge, use the same configuration for it.
The baselines differ from this setting as follows.
IntellAgent does not support the user-side tools of telecom, so we run it on the airline and retail agents, with as many valid conversations as \sys runs on them, and reuse the same scenarios with both underlying models.
We count a conversation as valid unless IntellAgent's simulated user leaks its reasoning into its messages, which exposes the intent of the test to the agent.
MANTRA runs its released suites with its own harness, in which the agent receives a single scenario message without a simulated user, and a prefix to the system prompt instructs it to complete the task without asking the user for confirmation or further input.
We therefore count neither a run that stops to ask the user nor one that acts without the confirmation that the policy requires as a bug.

\paragraph{Bug confirmation.}
An automated judge can report false violations, for example when a
specification is incorrect or the simulated user deviates from its
task. To compare approaches on the same footing, we manually review
every run of \sys and every violation a baseline reports. Two authors independently
examine each trace against the source specification and decide whether
the agent breaks it; disagreements are resolved by discussion. Only violations confirmed by this
review count as bugs.

\paragraph{Counting bugs.}
Many bugs $(s,\tau)$ (Definition~\ref{def:bug}) break the same refined specification: since a refined specification fixes its Given, When, and Then, such runs show the agent failing to enforce the same rule under the same conditions. We therefore deduplicate bugs by the refined specification they break and report \emph{unique bugs}, one per violated refined specification, however many runs violate it; hereafter, \emph{bug} refers to a unique bug. Deduplication also makes approaches comparable, since they run tests of different length and granularity and may expose the same bug repeatedly. A test of \sys counts a bug only when its run breaks the refined specification that it tests, a conservative choice that ignores violations of other specifications in the same run. For each violation of a baseline confirmed in the manual review, the reviewers identify the refined specifications whose Given holds in the run and whose Then it breaks.
We also exclude the \emph{turn rule}, which all three system prompts state: the agent must not send a message and call a tool in the same turn. The agents break it regardless of the specification under test, in 69.3\% and 32.2\% of the test cases with the default strategy on DeepSeek-V4-Flash and Qwen-3.8-Flash, so counting it would add the same violation to every approach.

\subsection{RQ1: Specification Extraction}
\label{sec:rq1}

\paragraph{Setup.}
Two authors independently annotate the correctness of the refined specifications. When they disagree, a third annotator with agent testing experience makes the final decision. We use Cohen's $\kappa$~\cite{cohen1960kappa} to measure the labeling disagreements.

\paragraph{Results.}

\begin{table}[t]
\centering
\caption{Specification counts by stage and quality of the refined specifications.}
\label{tab:rq1}
\small
\setlength{\tabcolsep}{8pt}
\begin{tabular}{@{}lrrrr@{}}
\toprule
Agent & Mined & Filtered & Refined & Correct \\
\midrule
Airline & 162 (65/41/56)  & 115 (55/24/36)  & 166 (89/24/53)  & 88.0\% \\
Retail  & 174 (73/43/58)  & 122 (47/20/55)  & 207 (98/20/89)  & 87.4\% \\
Telecom & 147 (76/25/46)  & 132 (62/24/46)  & 178 (74/24/80)  & 88.8\% \\
\midrule
\rowcolor{gray!15}
\textbf{Total} & \textbf{483 (214/109/160)} & \textbf{369 (164/68/137)} & \textbf{551 (261/68/222)} & \textbf{88.0\%} \\
\bottomrule
\end{tabular}

\par\smallskip
\begin{minipage}{0.9\linewidth}
\footnotesize
\textit{Note:} Parentheses show counts by source (system prompt/tool definitions/domain knowledge).
\end{minipage}
\end{table}

Table~\ref{tab:rq1} counts the specifications at each stage. Mining yields 483 specifications, and deduplication removes 120 of them. Binding repair adds 25 system-prompt specifications, and filtering removes 19 specifications inferred from domain knowledge, which leaves 369. Refinement turns them into 551 refined specifications, 1.5 per specification. The tool definitions and domain knowledge contribute 56\% of the filtered specifications. They state argument formats and preconditions that the system prompt leaves implicit, such as the requirement that a line belong to the customer who asks to change it, which underlies the ownership bugs of RQ4.

88.0\% of the refined specifications are correct ($\kappa = 0.88$): all of those from tool definitions, 91.2\% of those from the system prompt, and 80.6\% of those inferred from domain knowledge. The 66 incorrect refined specifications have three causes. 30 state rules without support in their source, such as refusing post-office-box addresses or forbidding roaming changes on a suspended line; 19 contradict the system prompt or the tool documentation, such as forbidding duplicate items that the exchange tool explicitly allows; and 17 drop a condition of the rule, such as letting the agent compensate a gold member who has not complained about a cancelled or delayed flight.

\answer{\textbf{Answer to RQ1:} 88.0\% of the refined specifications are correct, with most errors in rules inferred from domain knowledge.}

\subsection{RQ2: Effectiveness}
\label{sec:rq2}

\paragraph{Setup.}
For each agent under test, \sys runs three rounds of tests. The first round runs every refined specification that becomes a test case (504 of the 551) once as written, with the default user behavior (Table~\ref{tab:turns}). The next two rounds are strategy iterations (\S\ref{sec:guidance}): in each, every one of the 504 refined specifications receives one perturbation of its test, chosen with the bug experience memory that the earlier rounds update. We compare the bugs that each approach finds, their overlap, and the runs spent per bug.

\begin{figure}[tb]
\centering
\includegraphics[width=0.73\linewidth]{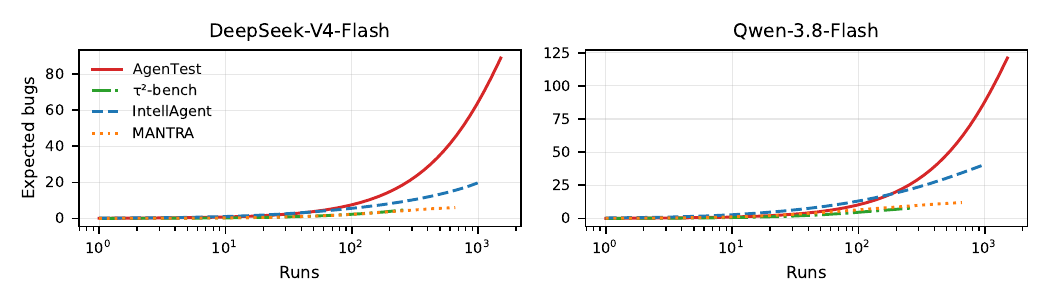}
\caption{Expected bugs found within a budget of runs.}
\label{fig:budget}
\includegraphics[width=0.73\linewidth]{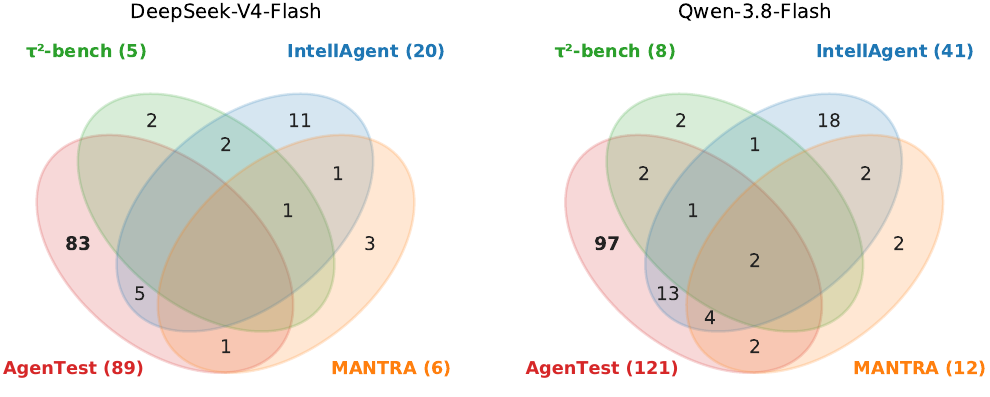}
\caption{Overlap of the bugs found by each approach, per underlying model. In parentheses: bugs per approach.}
\label{fig:venn}
\end{figure}

\paragraph{Results.}

\emph{Overall effectiveness.} Table~\ref{tab:rq2} shows that \sys finds the most bugs on every agent under test: 89 bugs on DeepSeek-V4-Flash and 121 on Qwen-3.8-Flash, 3.4 and 2.5 times the 26 and 49 bugs that the three baselines find together. On the airline and retail agents, which all four approaches test, \sys finds 64 and 89 bugs against 20 and 41 for the strongest baseline, IntellAgent (3.2 and 2.2 times). It also needs the fewest runs per bug with both underlying models, 17.0 and 12.5 against 51.0 and 24.9 for IntellAgent.

\begin{wraptable}{r}{0.48\textwidth}
\centering
\caption{Bugs (violated refined specifications) per approach. Runs/bug: runs per bug, counting tasks for $\tau^2$-bench, single-turn runs for MANTRA, and conversations otherwise.}
\label{tab:rq2}

\small
\setlength{\tabcolsep}{4pt}

\begin{tabular}{@{}lrrrrr@{}}
\toprule
\multicolumn{6}{c}{DeepSeek-V4-Flash} \\
\cmidrule(lr){1-6}
Approach & Air. & Ret. & Tel. & Total & Runs/bug \\
\midrule
$\tau^2$-bench & 3 & 1 & 1 & 5 & 55.6 \\
IntellAgent & 13 & 7 & -- & 20 & 51.0 \\
MANTRA & 2 & 1 & 3 & 6 & 110.0 \\
\midrule
\rowcolor{gray!15}
\sys & \textbf{23} & \textbf{41} & \textbf{25} & \textbf{89} & \textbf{17.0} \\
\midrule
\multicolumn{6}{c}{Qwen-3.8-Flash} \\
\cmidrule(lr){1-6}
Approach & Air. & Ret. & Tel. & Total & Runs/bug \\
\midrule
$\tau^2$-bench & 0 & 5 & 3 & 8 & 34.8 \\
IntellAgent & 19 & 22 & -- & 41 & 24.9 \\
MANTRA & 2 & 8 & 2 & 12 & 55.0 \\
\midrule
\rowcolor{gray!15}
\sys & \textbf{37} & \textbf{52} & \textbf{32} & \textbf{121} & \textbf{12.5} \\
\bottomrule
\end{tabular}

\end{wraptable}

\emph{Under equal budget.} The approaches run different numbers of tests, so Figure.~\ref{fig:budget} compares them at equal budgets.
For a budget of $b$ runs, the curve shows the average number of bugs exposed by $b$ randomly sampled runs of an approach.
Averaging over random picks makes the comparison independent of the order in which each approach happened to run its tests. \sys leads at every budget beyond 44 runs on DeepSeek-V4-Flash and 178 on Qwen-3.8-Flash. Below that, IntellAgent finds slightly more, because its few bugs recur in many conversations: its most frequent one appears in 41 and 93 of them. Within the 278 runs of $\tau^2$-bench, \sys finds 20.3 bugs on average against 5 on DeepSeek-V4-Flash, and 27.7 against 8 on Qwen-3.8-Flash; within the 660 runs of MANTRA, it finds 45.2 and 61.7 against 6 and 12.

\emph{Overlap and complementarity.} \sys finds 83 bugs on DeepSeek-V4-Flash and 97 on Qwen-3.8-Flash that no baseline finds, out of its 89 and 121 (Figure~\ref{fig:venn}). Conversely, the three baselines together find 26 and 49 bugs, of which \sys finds 6 and 24. Every bug that only a baseline finds breaks a refined specification that \sys tests without exposing a violation. Most of them arise in long tasks that chain several goals, such as troubleshooting several faults before a transfer, or booking a flight and asking for compensation in the same conversation. \sys tests these refined specifications one at a time and does not observe these bugs: a test of a single specification does not create the interactions between goals that a long task does.

\emph{Why the baselines miss these bugs.} The baselines build their tests from tasks rather than from a model of the specifications that the agent must satisfy, so nothing steers their tests toward specifications that remain untested. Their tests keep hitting the same few specifications: on average, a bug of IntellAgent is exposed by 5.8 and 8.4 conversations, and a bug of MANTRA by 3.0 and 6.9 runs, against 1.3 and 1.3 runs for \sys, whose tests each target one refined specification. For IntellAgent, 70 of the 93 conversations with a confirmed violation on DeepSeek-V4-Flash and 169 of 280 on Qwen-3.8-Flash break a single rule, that the agent must not state information that neither the user nor a tool provided. Under the same budget, such repeated hits add no new bug. The specification model also tells \sys which states to construct, such as a reservation that belongs to another user to test an ownership check, which task-based tests seldom set up. $\tau^2$-bench, in addition, judges a run mainly by the final database state, which misses violations that leave the state unchanged, such as acting before the user confirms.

\answer{\textbf{Answer to RQ2:} \sys finds 3.4 and 2.5 times as many bugs as all baselines combined (89 vs.\ 26 and 121 vs.\ 49), and leads at every equal budget beyond 178 runs. It finds 83 and 97 bugs that no baseline finds.}
\subsection{RQ3: Component Effectiveness}
\label{sec:rq3}

We evaluate each component of specification-driven testing (\S\ref{sec:testing}) on DeepSeek-V4-Flash, under the three-round setting of RQ2.

\begin{wraptable}{r}{0.5\linewidth}
\vspace{-0.6\baselineskip}
\centering
\caption{Test driver construction and hybrid bug detection on DeepSeek-V4-Flash. Default: round~1 (R1), with the default strategy; Library: rounds 2--3 (R2--3), with strategies from the library.}
\label{tab:rq3}
\footnotesize
\setlength{\tabcolsep}{3pt}
\begin{tabular}{@{}lrr@{}}
\toprule
Measure & R1 & R2--3 \\
\midrule
Drivers constructed & 504/551 & -- \\
Runs that trigger \emph{When} as \emph{Given} requires   & 500/504 & 1003/1008 \\
\bottomrule
\end{tabular}
\vspace{-0.8\baselineskip}
\end{wraptable}

\paragraph{Test Driver Construction}
\label{sec:rq3-driver}
To evaluate the test driver, we check whether it can be constructed for each refined specification and whether, in each run, the simulated user states the user-side conditions of Given and triggers When.
Table~\ref{tab:rq3} shows that \sys constructs drivers for 504 of the 551 refined specifications (91.5\%). The other 47 fail because 26 specifications are incorrect (RQ1) and 21 require a state that the environment cannot hold, such as a ported-out line.
In 1,503 of the 1,512 runs (99.4\%), the simulated user states these conditions and triggers When. Of the other nine, four runs with the default strategy end early because the user cannot supply what the test needs, such as a new address or the result of a device check, and five runs with a library strategy the agent itself refuses the transfer that the refined specification presupposes.
These results show that the driver effectively establishes the preconditions of the refined specifications.

\paragraph{Adaptive Test Case Generation}
\label{sec:rq3-generation}
To evaluate adaptive test case generation, we compare \sys with two variants: one that uses only the default strategy and one that selects strategies at random instead of with the bug experience memory.
Table~\ref{tab:ablation} shows that the default strategy alone violates only 6 refined specifications, against 89 for \sys; random selection violates 84.
These results show that the interaction strategies account for most of the bugs, and that the bug experience memory finds further bugs on top of them.

\paragraph{Hybrid Bug Detection}
\label{sec:rq3-detection}
To evaluate hybrid bug detection, we compare the verdicts of the oracle against the manual review described in \S\ref{sec:setup}, which serves as ground truth.
Based on this comparison, we measure two metrics: the agreement rate, the proportion of runs on which the verdict of the oracle matches the review, and the precision, the proportion of reported violations that the review confirms.
Unlike the bug rate in Table~\ref{tab:ablation}, precision also counts violations of the turn rule and of excluded specifications.
For comparison, we measure the precision of the judge of each baseline in the same way.
As shown in Table~\ref{tab:precision}, the oracle agrees with the manual review on 1,476 of the 1,512 runs (97.6\%) and achieves a precision of 82.6\% (161 of 195 reported violations).
In contrast, the judges of $\tau^2$-bench, IntellAgent, and MANTRA achieve precisions of only 25.0\%, 46.5\%, and 5.7\%, respectively (Table~\ref{tab:precision}).
The precision of \sys further increases to 90.4\% (142 of 157) in the runs with library strategies, which expose most of the bugs.
Moreover, in the test cases with the default strategy, 492 of the 579 checks (85\%) are deterministic rules on tool calls and environment states, so most verdicts are independent of LLM judgment.
These results demonstrate that the hybrid oracle detects violations accurately and with substantially higher precision than the judges of the baselines.

\answer{\textbf{Answer to RQ3:} The test driver is constructed for 91.5\% of the refined specifications and triggers When in 99.4\% of the runs. Interaction strategies account for most of the bugs (89 vs. 6), and the bug experience memory adds 5 more. The hybrid oracle achieves a precision of 82.6\%, against at most 46.5\% for the baselines.}

\begin{table}[t]
\small
\begin{minipage}[t]{0.53\linewidth}
\centering
\caption{Adaptive test generation and execution on DeepSeek-V4-Flash. Bug rate: share of runs with a bug.}
\label{tab:ablation}
\setlength{\tabcolsep}{4pt}
\begin{tabularx}{\linewidth}{@{}>{\raggedright\arraybackslash}Xrrr@{}}
\toprule
Variant &
\begin{tabular}[b]{@{}r@{}}Violated\\specs\end{tabular} &
Runs &
\begin{tabular}[b]{@{}r@{}}Bug\\rate\end{tabular} \\
\midrule
Default strategy only   & 6           & 504   & 1.2\% \\
Random strategy selection    & 84          & 1,512 & 6.9\% \\
\sys                             & \textbf{89} & 1,512 & \textbf{7.7\%} \\
\bottomrule
\end{tabularx}
\end{minipage}%
\hfill%
\begin{minipage}[t]{0.43\linewidth}
\centering
\caption{Failures that each approach's own judge reports (Rep.) and manual review confirms (Conf.), and their ratio (Prec.). }
\label{tab:precision}
\setlength{\tabcolsep}{4pt}
\begin{tabular}{@{}lrrr@{}}
\toprule
Approach       & Rep. & Conf. & Prec. \\
\midrule
$\tau^2$-bench & 28   & 7     & 25.0\% \\
IntellAgent    & 200  & 93    & 46.5\% \\
MANTRA         & 281  & 16    & 5.7\% \\
\sys           & 195  & 161   & \textbf{82.6\%} \\
\bottomrule
\end{tabular}
\end{minipage}
\end{table}

\subsection{RQ4: Bug Analysis}
\label{sec:rq4}

We analyze the business logic bugs that \sys finds to understand how agents violate business rules and to guide their repair and future testing.
To this end, we classify each bug along two dimensions.
The \emph{symptom} describes what the agent does wrong and is determined by the Then of the violated specification, which yields six categories.
The \emph{root cause} describes why the agent does it and is determined by two questions: whether the rule is stated in the system prompt or the tool definitions, and if so, whether the agent attempts to follow it.
This yields three categories: the rule is \emph{implicit}, the stated rule is \emph{ignored}, or the stated rule is \emph{misapplied}.
 For each run with a bug, we also check whether the bug is visible in the final state, that is, whether the run changes the database to a state that no compliant run could produce. Two authors independently label the symptom and the root cause of each bug. When they disagree, a third annotator with agent testing experience makes the final decision. Over all labels, Cohen's $\kappa$~\cite{cohen1960kappa} is 0.93.

\begin{table}[t]
\caption{Bugs found by \sys by symptom and root cause.}
\label{tab:symptoms}
\scriptsize
\setlength{\tabcolsep}{3pt}
\begin{tabularx}{\linewidth}{@{}l>{\raggedright\arraybackslash}Xcrrrrrrrr@{}}
\toprule
\multirow{2}{*}{Symptom} & \multirow{2}{*}{Description} & \multirow{2}{*}{Agents}
 & \multicolumn{2}{c}{Implicit} & \multicolumn{2}{c}{Ignored} & \multicolumn{2}{c}{Misapplied} & \multicolumn{2}{c}{Total} \\
\cmidrule(lr){4-5}\cmidrule(lr){6-7}\cmidrule(lr){8-9}\cmidrule(l){10-11}
 & & & DS & Qwen & DS & Qwen & DS & Qwen & DS & Qwen \\
\midrule
Unauthorized action     & Acts on another user's object                        & A, T    & 8  & 9  & 1  & 1  & 0  & 0  & 9  & 10 \\
Disallowed operation    & Acts despite a forbidding state, condition, or limit & A, R, T & 4  & 5  & 5  & 2  & 4  & 15 & 13 & 22 \\
Missing prerequisite    & Acts before a required confirmation or check        & A, R, T & 0  & 0  & 7  & 15 & 5  & 10 & 12 & 25 \\
Invalid argument        & Uses an invalid or unsourced argument value          & A, R, T & 31 & 28 & 11 & 14 & 1  & 4  & 43 & 46 \\
Improper escalation     & Transfers at the wrong time, or acts after it        & R, T    & 2  & 1  & 0  & 0  & 4  & 9  & 6  & 10 \\
Incorrect communication & States unsupported facts or omits disclosures & A, R, T & 0  & 1  & 3  & 0  & 3  & 7  & 6  & 8  \\
\midrule
Total                   &                                                      &         & 45 & 44 & 27 & 32 & 17 & 45 & \textbf{89} & \textbf{121} \\
\bottomrule
\end{tabularx}

\par\smallskip
\begin{minipage}{\linewidth}
\scriptsize
\textit{Note:} DS: DeepSeek-V4-Flash; Qwen: Qwen-3.8-Flash; A/R/T: airline/retail/telecom.
\end{minipage}

\end{table}

Table~\ref{tab:symptoms} presents the bugs by symptom and root cause.
Overall, the symptoms are consistent across the two models: invalid arguments are the most common symptom, and most bugs concern tool calls on business data.
Specifically, invalid arguments account for 43 of the 89 bugs on DeepSeek-V4-Flash (48.3\%) and 46 of the 121 on Qwen-3.8-Flash (38.0\%), followed by disallowed operations (13 and 22) and missing prerequisites (12 and 25), whereas unauthorized actions (9 and 10), improper escalation (6 and 10), and incorrect communication (6 and 8) are less frequent.
Together, the four symptoms that concern tool calls on business data, namely invalid arguments, disallowed operations, missing prerequisites, and unauthorized actions, account for 86.5\% and 85.1\% of the bugs.
In terms of agents, four of the six symptoms occur in all three agents.
In terms of observability, 60\% and 59\% of the runs with a bug cannot be detected from the final state alone: 42\% and 37\% leave the database unchanged, and the others write a state that a compliant run could also produce, such as acting on another user's reservation or acting without an explicit yes.
These results indicate that business logic bugs manifest mainly in what the agent does to business data rather than in what it says to the user, that they recur across domains rather than being specific to a single agent, and that most of them can be detected only by checking the full interaction against the business rules.

From the perspective of root causes, all three root causes account for a substantial share of the bugs on both models.
Specifically, implicit, ignored, and misapplied rules account for 50.6\%, 30.3\%, and 19.1\% of the bugs on DeepSeek-V4-Flash, and for 36.4\%, 26.4\%, and 37.2\% on Qwen-3.8-Flash.
In terms of symptoms, unauthorized actions arise almost entirely from implicit rules (8 of 9 and 9 of 10), whereas missing prerequisites arise only from stated rules.
The two models differ mainly in misapplied rules, which are far more frequent on Qwen-3.8-Flash (45 vs.\ 17).
These results suggest that business logic bugs stem both from rules that no artifact states and from stated rules that the agent fails to follow, and that each root cause calls for a different repair: implicit rules must first be stated in the system prompt or enforced by the tools, ignored rules call for enforcement in the tools, and misapplied rules call for a clearer statement of the rule.

\answer{\textbf{Answer to RQ4:} Most bugs manifest in tool calls on business data (86.5\% and 85.1\%), with invalid arguments the most common symptom, and about 60\% of the runs with a bug cannot be detected from the final state alone. They stem from implicit rules as well as from stated rules that the agent ignores or misapplies.}

\section{Discussion}

\textit{Implicit rules expose gaps in agent artifacts.}
Half of the bugs on DeepSeek-V4-Flash (45 of 89) and over a third on Qwen-3.8-Flash (44 of 121) break a business rule that neither the system prompt nor the tool definitions state (Table~\ref{tab:symptoms}).
For such a bug, the agent cannot be expected to follow a rule it is never given, yet the business would not accept the behavior, as when the agent changes another user's reservation.
We therefore regard a bug of an implicit rule as a defect of the agent as a whole, whose repair may lie in the system prompt or the tools rather than in the underlying model.

\textit{Structured rules should be enforced in the tools.}
About 60\% of the runs with a bug cannot be detected from the final state alone, and unauthorized actions succeed because the tools accept any user's object (\S\ref{sec:rq4}).
Rules with structured evidence, such as ownership, argument values, and call order, need not be left to the model: 85\% of the oracle checks in the default test cases are deterministic predicates over tool calls and states, which could also guard tool execution at runtime~\cite{kamath2025agentc}.
Conversational rules, such as confirmation, remain with the model, and testing remains their main safeguard.

\textit{Some bugs emerge only when goals are chained.}
Every bug that only a baseline finds breaks a refined specification that \sys also tests, but arises in a task that chains several goals in one conversation (\S\ref{sec:rq2}).
This suggests that goals sharing a conversation can interfere with how an agent follows a rule.
For agent design, decomposing a user request into single-goal subtasks may reduce such interference, provided that rules spanning subtasks, such as a limit on compensation per incident, are enforced across them.
For testing, \sys deliberately tests one refined specification at a time, so that each violation is attributable to a single rule; composing refined specifications that share tools into multi-goal tests would complement it.

\textit{Bugs depend on the underlying model.}
With the same system prompt, tools, and refined specifications, the two underlying models differ both in how many rules they break and in why.
\sys finds 89 bugs on DeepSeek-V4-Flash and 121 on Qwen-3.8-Flash, of which misapplied rules account for 17 and 45 (Table~\ref{tab:symptoms}), and the agents break the turn rule in 69.3\% and 32.2\% of the test cases with the default strategy (\S\ref{sec:setup}).
The testing results of one model therefore do not carry over to another.
Since refined specifications depend only on the system prompt and the tool definitions, they can be mined once and rerun as a regression suite whenever the underlying model is upgraded or replaced.

\section{Threats to Validity}
\label{sec:threats}

\paragraph{Internal validity.}
Errors in extraction, user simulation, and LLM judgment may affect our findings; we mitigate them with assertions fixed before execution, evidence checked against the trace, a judge model distinct from the agents' models, and two-author review of every run of \sys and every reported baseline violation. Results may vary across runs, though they agree across both underlying models.

\paragraph{Construct validity.}
Counting bugs by refined specification depends on how \sys mines and refines specifications. We apply it to all approaches, and every confirmed baseline violation maps to a refined specification, though incomplete specifications may limit what \sys explores.

\paragraph{External validity.}
Our evaluation covers three domains of one benchmark and two underlying models, and may not generalize to other agents or deployments. \sys requires the system prompt, the tool definitions, and a resettable environment with a known data model, and its interaction strategies remain untested in unseen domains.

\section{Related Work}
\label{sec:related}

\paragraph{Testing of Tool-Using Agents.}
Existing work spans three categories.
Benchmark-based evaluation uses predefined tasks~\cite{yao2024taubench,barres2025tau2,zhou2024webarena,drouin2024workarena,liu2024agentbench} or tasks synthesized from procedural documents~\cite{mantra}.
Much of this benchmark-based evaluation focuses on safety, particularly harmful actions and prompt injection~\cite{toolemu,agentdojo,injecagent,agentharm}. Automated testing generates policy-based scenarios, exercises mined workflow boundaries, applies coverage-guided fuzzing, or uses adversarial conversations~\cite{levi2025intellagent,komoravolu2026ata,flare,lin2026agenteval,nakash2025craft}.
Other approaches assess recorded traces against prompt-derived rules~\cite{sharma2026agentpex} or enforce temporal constraints at runtime~\cite{kamath2025agentc}. None of them makes each business specification an explicit test target, so implicit specifications and those that need a specific state are rarely exercised.
In contrast, \sys mines each specification, constructs a scenario in which it applies, and checks the agent's behavior against it.

\paragraph{Specification-Driven Testing.}
Specification-driven testing uses formal models to generate tests and check conformance~\cite{hierons2009using}, natural-language requirements and behavioral scenarios to guide testing~\cite{north2006bdd,yang2026sega}, and protocol knowledge to guide message and sequence generation in fuzzing~\cite{meng2024chatafl}. These approaches rely on explicit requirements, behavioral models, or protocol specifications. Tool-using agents often lack comparable standalone business specifications: their rules are distributed across system prompts and tool definitions, with some left implicit. \sys addresses this gap by making these rules explicit and using them to coordinate environment setup, multi-turn interaction, and violation detection.

\section{Conclusion}

We presented \sys, which mines business specifications from agent artifacts and domain knowledge and uses each one to drive environment setup, multi-turn interaction, and violation detection. On three domains of $\tau^2$-bench, \sys detects violations of 89 and 121 distinct specifications on two underlying models, exposing recurring failures in authorization, confirmation, and workflow enforcement that task completion alone does not reveal. Future work will improve the validation of domain-inferred specifications and test interactions among multiple specifications.

\section*{Data Availability}

To support reproducibility, we release the source code of \sys at \url{https://anonymous.4open.science/r/VIOLET-D190}

\bibliographystyle{ACM-Reference-Format}
\bibliography{sample-base}

\end{document}